\documentclass[12pt]{article}
\usepackage{amsmath,bbm,amssymb}
\usepackage{amsthm}
\usepackage{comment}
\usepackage{graphicx, graphics}
\usepackage{textcomp}
\usepackage{amsfonts}

\usepackage{psfrag,pst-node,subfigure,rotating, amsmath, bbm, amsthm, amssymb, amsthm, setspace, picture, epsfig, amsfonts, upgreek}

\usepackage{multirow}

\usepackage{arydshln}
\usepackage{algorithm}
\usepackage{algorithmic}

\UseRawInputEncoding  % seems a key to avoid error in arxiv
\usepackage{arydshln}

\newtheorem{theorem}{Theorem}%[section]
\def\v{\boldsymbol}\def\m{\mathbf}
\def\s{\mathcal}
\begin{document}
\begin{center}
    {\Large\bf Discretization approximation: An alternative to Monte Carlo in Bayesian computation}
\\[2mm] {\large Shifeng Xiong$^{1~2}$}
\\ 1. SKLMS, Academy of Mathematics and Systems Science, Chinese Academy of Sciences, Beijing 100190, China\\
2. School of Mathematical Sciences, University of Chinese Academy of Sciences,\\ Beijing 100049, China\\xiong@amss.ac.cn
\end{center}

\vspace{1cm} \noindent{\bf Abstract}\quad In this paper we propose a new deterministic approximation method, called discretization approximation, for Bayesian computation. Discretization approximation is very simple to understand and to implement. It only requires calculating posterior density values as probability masses at pre-specified support points. The resulting discrete distribution can be a good approximation to the target posterior distribution. All posterior quantities, including means, standard deviations, and quantiles, can be approximated by those of this completely known discrete distribution. We establish the convergence rate of discretization approximation as the number of support points goes to infinity. If the support points are generated from quasi-Monte Carlo sequences, then the rate is actually the same as that in integration approximation, generally faster than the optimal statistical rate. In this sense, discretization approximation is superior to the popular Markov chain Monte Carlo method. We also provide random sampling and representation point construction methods based on discretization approximation. Numerical examples including some benchmarks demonstrate that the proposed method performs quite well for both low-dimensional and high-dimensional cases.

%\vspace{1cm} \noindent{\bf AMS 2000 subject classifications:} 62A01, 62C05.

\vspace{1cm} \noindent{{\bf KEY WORDS:} Markov chain Monte Carlo; proposal distribution; quasi-Monte Carlo; representation point; space-filling design.}

%\vspace{1cm} \noindent{{\bf Mathematics Subject Classification:} 65C60, 65C05, 62F15}

%\newpage

\section{Introduction}\label{sec:intro}

In this paper we discuss computing quantities of a multivariate distribution when only the density function, up to a normalizing constant, is available. This is a fundamental problem in modern statistics, especially in Bayesian statistics, which frames all inferences on unknown parameters as a calculation about the posterior density.
Nowadays Monte Carlo (MC) methods, implemented mainly with Markov chain Monte Carlo (MCMC) techniques, have become the first choice for most problems (Robert and Casella 2004; Gelman, Carlin, Stern et al. 2013). These methods suffer less from the curse of dimensionality and can obtain the results with arbitrary precision via persistent sampling. MCMC techniques have greatly promoted applications of Bayesian methods. As a result, the Metropolis-Hastings algorithm (Metropolis, Rosenbluth, Rosenbluth et al. 1953; Hastings 1970), one of the main MCMC techniques, was ranked among the top ten algorithms of the 20th century scientific computing (Cipra 2000).

However, in real applications MCMC may have high computational cost for complex models, as well as other technical problems such as convergence.
To overcome these difficulties, other approximation methods have been proposed in the literature. They use some simplified distribution to approximate the target distribution, and then approximate the posterior quantities by those of the simplified distribution. Such simplifications include the Laplace approximation method (Bornkamp 2011; Rue, Martino, and Chopin 2009), expectation propagation (Minka 2001), variational inference (Blei, Kucukelbir, and McAuliffe 2017), discretization with mixture of uniform distributions (Fu and Wang 2002; Wang and Lee 2014), and approximations via Gaussian process interpolation (Joseph 2012; Conrad, Marzouk, Pillai et al. 2016), among many others. These methods are effective for specific cases, yet may have their own problems in uncontrollable bias, complex implementation, or less theoretical guarantee.

In this paper we propose another approximation method, called discretization approximation, which is based on an straightforward discretization to the density function. Discretization approximation is very simple to understand and to implement. It only requires calculating density values as probability masses at pre-specified support points. The resulting discrete distribution can be a good approximation to the target distribution. All quantities, including means, standard deviations, and quantiles, can be approximated by those of this completely known discrete distribution, which are easy to obtain.

We establish the convergence rate of the discrete distribution to the target distribution as the number $M$ of support points goes to infinity.
If the support points are generated from quasi-Monte Carlo sequences (Niederreiter 1992), then the rate is actually the same as that in integration approximation, generally faster than the optimal statistical rate $O_p\left(1/\sqrt{M}\right)$. In this sense, discretization approximation is superior to all MC methods including the popular MCMC method.
If the support points are generated from random sampling, then convergence rate of discretization approximation is independent of the dimension like Monte Carlo. In fact, modern quasi-Monte Carlo rules can also have a convergence rate independent of the dimension (Joe 2004; Dick, Kuo, and Sloan 2013). These theoretical properties guarantee that discretization approximation can perform well in both low-dimensional and high-dimensional problems.
Furthermore, discretization approximation is a deterministic procedure, and thus does not involve any uncertainty from random sampling. From the computational perspective, discretization approximation possesses a low complexity, requiring only $O(M)$ operations.

Discretization approximation also provides an approximate sampling method from the target distribution. Clearly, a sample generated from the above discrete distribution can be viewed as an approximate one from the desired distribution. We present both random sampling and deterministic design strategies to implement this approximate sampling. The former possesses the optimal statistical convergence rate, and the latter can deterministically converge faster than that optimal rate. With better representativeness, the points from the deterministic procedure can act as representation points of the target distribution.

Numerical examples including some benchmarks in this area are used to evaluate discretization approximation. The results are encouraging, demonstrating that the proposed method performs well for all these examples in both low-dimensional and high-dimensional cases. In particular, it outperforms popular MC/MCMC methods in terms of accuracy and/or speed.

This paper is organized as follows. Section \ref{sec:method} introduces discretization approximation for the target distribution whose support set is the unit hypercube. Theoretical properties on population approximation and the corresponding sampling procedure are presented. Section \ref{sec:unb} discusses discretization approximation for general distributions, and provides the implementation details. Section \ref{sec:ne} presents numerical comparisons between the proposed method and existing methods. We conclude the paper with some discussion in Section \ref{sec:dis}. Technical proofs are given in the Appendix.

\section{Approximation on the unit hypercube}\label{sec:method}

\subsection{Population approximation}\label{subsec:dc}
We consider the problem of approximating a target distribution whose probability density function $f$ is available up to an (unknown) constant. Let $\s{S}\subset{\mathbb{R}}^d$  denote the support of $f$. In this section we assume that $\s{S}$ is bounded. Without loss of generality, let $\s{S}\subset[0,1]^d$. Write $\mu_f$ for the probability measure on $[0,1]^d$ corresponding to $f$.

\begin{figure}[t]\begin{center}
\scalebox{0.55}[0.55]{\includegraphics{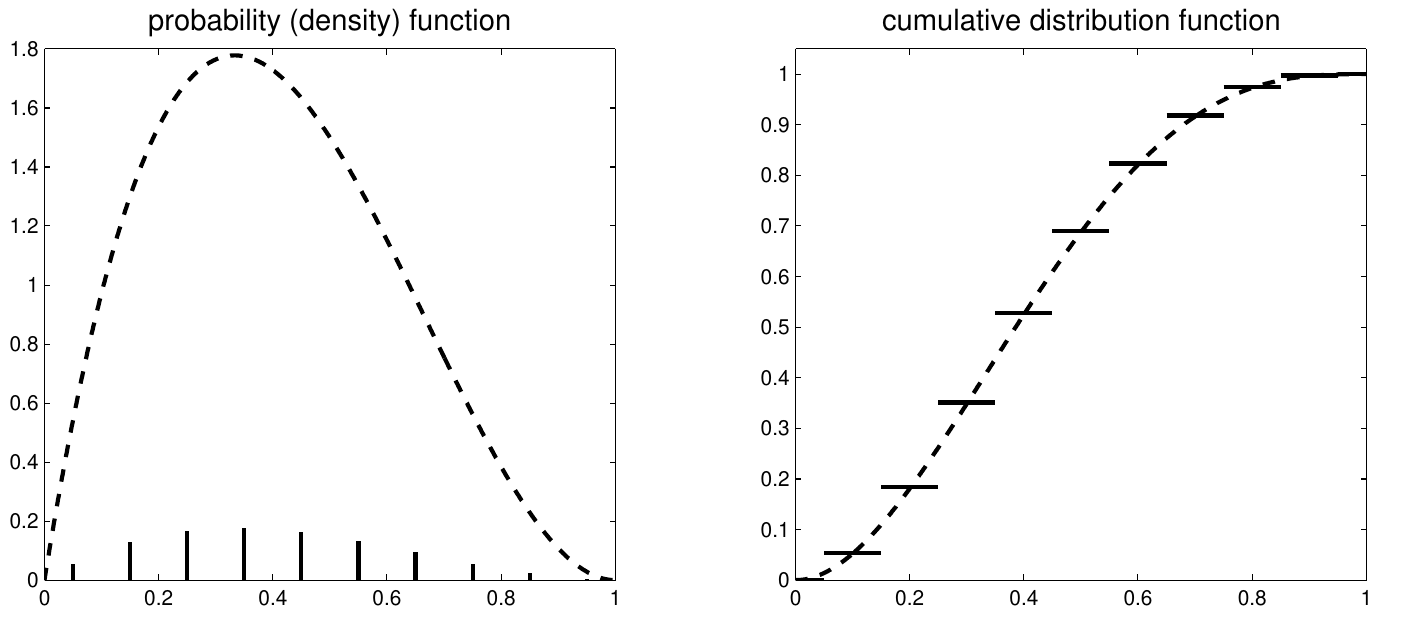}}
\end{center}
\caption{Comparisons between $\mu_{(f,\s{A}_M)}$ (the solid lines) and $\mu_f$ (the dashed lines), where $f$ is the density of the beta distribution $Beta(2,3)$ and $\s{A}_M=\{(2i-1)/(2M)\}_{i=1}^{M}$ with $M=10$. }\label{fig:beta}
\end{figure}

The basic idea of our method is discretization of $f$. First take a support point set $\s{A}_M=\{\v{a}_i\}_{i=1}^M\subset[0,1]^d$ for a large integer $M$. The elements of $\s{A}_M$ are required to be uniformly scattered. Define a probability measure $\mu_{(f,\s{A}_M)}$ on $\s{A}_M$ as \begin{equation}\label{dp}P(\v{X}_{(f,\s{A}_M)}=\v{a}_i)=p_i=\frac{f(\v{a}_i)}{\sum_{j=1}^Mf(\v{a}_j)},\ \ i=1,\ldots,M,\end{equation}
where the random vector $\v{X}_{(f,\s{A}_M)}\sim\mu_{(f,\s{A}_M)}$. Note that the discrete measure $\mu_{(f,\s{A}_M)}$ is completely known. It can be expected that, if $\s{A}_M$ possesses some dense properties on $[0,1]^d$, then $\mu_{(f,\s{A}_M)}$ will become very close to $\mu_f$ as $M$ increases. Hence, all quantities of $\mu_f$, including the means, quantiles, and standard deviations, can be approximated by those of $\mu_{(f,\s{A}_M)}$. We refer to this method as discretization approximation. Figure \ref{fig:beta} presents a simple example that compares $\mu_{(f,\s{A}_M)}$ and $\mu_f$ when $f$ is the density of a Beta distribution. It can be seen that the approximation is quite effective even with only $M=10$.

A feasible way to generate the support point set $\s{A}_M$ is random sampling from the uniform distribution on $[0,1]^d$. Deterministic space-filling designs are usually more effective (Santner, Williams, and Notz 2018). Such designs include minimax distance designs (Johnson, Moore, and Ylvisaker 1990), quasi-Monte Carlo sequences (Niederreiter 1992), optimal Latin-hypercube designs (Park 2001), and many others. Here we adopt the quasi-Monte Carlo methods. The first reason is that there are a number of mature procedures to construct them. The second reason, maybe more importantly, is that they give deterministic error bounds of integration approximation, which can be used to derive the error bounds of discretization approximation to the target distribution. Let $\s{A}_M=\{\v{a}_i\}_{i=1}^M$ be a quasi-Monte Carlo sequence satisfying the error bound,
\begin{equation}\label{eb} \left|\frac{s(\v{a}_1)+\cdots+s(\v{a}_M)}{M}-\int_{[0,1]^d}s(\v{x})\mathrm{d}\v{x}\right|\leqslant \delta_{M,d}V(s) \end{equation}
for any integrable function $s$ and sufficiently large $M$, where $\delta_{M,d}$ is a positive number dependent only on $M$ and $d$, $\delta_{M,d}\to0$ as $M\to\infty$ for fixed $d$, and $V(s)$ is the variation of $s$ in the sense of
Hardy and Krause; see for instance Niederreiter (1992). In this paper we assume $V(f)<\infty$. The bound $\delta_{M,d}$, also known as the discrepancy of the point set $\s{A}_M=\{\v{a}_i\}_{i=1}^M$ (Fang, Li, and Sudjianto 2006), represents the convergence rate of the integration approximation with a quasi-Monte Carlo sequence. For example, the Halton sequence
gives $\delta_{M,d}=C_d(\log M)^d/M$, where $C_d$ depends only on $d$ (Dick, Kuo, and Sloan 2013). It is worth pointing out that $\delta_{M,d}$ corresponds to the worst-case error. For a sufficiently smooth function, the actual convergence rate can be much faster.

To discuss theoretical properties of discretization approximation, we need some notation and definitions. Write $\v{x}'$ for the transpose of vector $\v{x}$. Let $d_\mathrm{K}(\nu_1,\ \nu_2)=\\\sup_{\v{x}\in[0,1]^d}\left|\nu_1\left([\v{0},\v{x}]\right)-\nu_2\left([\v{0},\v{x}]\right)\right|$ denote the Kolmogorov distance between two probability measures $\nu_1$ and $\nu_2$ on $[0,1]^d$, where $[\v{0},\v{x}]=[0,x_1]\times\cdots\times[0,x_d]$ for $\v{x}=(x_1,\ldots,x_d)'$. For $d$ nonnegative integers $k_1,\ldots,k_d$ and a probability measure $\nu$, let $E_{(k_1,\ldots,k_d)}(\nu)=E\left(X_1^{k_1}\cdots X_d^{k_d}\right)$ with $(X_1,\ldots,X_d)'\sim \nu$. Denote $f_{(k_1,\ldots,k_d)}(\v{x})=x_1^{k_1}\cdots x_d^{k_d}f(\v{x})$ for $\v{x}=(x_1,\ldots,x_d)'\in[0,1]^d$.

\begin{theorem}\label{th:wc} For sufficiently large $M$, we have \begin{equation}\label{feb} d_\mathrm{K}(\mu_{(f,\s{A}_M)},\ \mu_f)\leqslant C\delta_{M,d},\end{equation}where $C$ is a positive constant independent of $M$ and $d$. Furthermore, if $E_{(k_1,\ldots,k_d)}(\mu_f)<\infty$ and $V(f_{(k_1,\ldots,k_d)})<\infty$, then for sufficiently large $M$,\begin{equation}\label{jfeb} \left|E_{(k_1,\ldots,k_d)}(\mu_{(f,\s{A}_M)})-E_{(k_1,\ldots,k_d)}(\mu_f)\right|\leqslant \tilde{C}\delta_{M,d},\end{equation} where $\tilde{C}$ is a positive constant independent of $M$ and $d$.\end{theorem}

The condition $V(f_{(k_1,\ldots,k_d)})<\infty$ for \eqref{jfeb} holds if $f$ possesses some smoothness properties such as partial derivatives' continuity (Niederreiter 1992).

In Bayesian statistics, all inferences are based on the posterior density $f$. Theorem \ref{th:wc} indicates that the completely known distribution $\mu_{(f,\s{A}_M)}$ are very close to $\mu_f$, and that the approximation error can be arbitrarily small. Therefore, the means, quantiles, and standard deviations of $\mu_{(f,\s{A}_M)}$,  which are easy to compute, can be used to approximate those of $\mu_f$. For example, the mean vector of $\mu_f$ can be computed as $\sum_{i=1}^Mp_i\v{a}_i$, where $p_i$ is given by \eqref{dp}. This approximation is superior to the sampling methods such as the MCMC methods because its convergence rate $\delta_{M,d}$ in Theorem \ref{th:wc} is usually much faster than the optimal statistical rate $O_p(1/\sqrt{M})$, achieved by Monte Carlo, with sample size $M$. As mentioned above, the Halton sequence gives $\delta_{M,d}=C_d(\log M)^d/M$. For smooth $f$, such a rate can be much higher. Assume that $f$ has smoothness of order $\alpha$, where $\alpha$ is an arbitrary integer. The quasi-Monte Carlo methods with higher-order sequences/digital nets give $\delta_{M,d}=C_d(\log{M})^{\alpha d}/M^\alpha$ (Dick 2009).

It is well known that the convergence rate $O_p(1/\sqrt{M})$ of Monte Carlo is independent of the dimension. If the support points are generated by Monte Carlo, then the proposed discretization approximation method also possesses this appealing property. Because the variances of these Monte Carlo methods rely on the function $f$ that involves the dimension, the real approximation errors actually depend on the dimension. In high-dimensional cases, modern quasi-Monte Carlo rules can have a prescribed rate of convergence for sufficiently smooth functions, and ideally also guaranteed slow growth (or no growth) of the worst-case error as $d$ increases (Joe 2004; Dick, Kuo, and Sloan 2013). Therefore, even for high dimensions, we recommend using quasi-Monte Carlo in discretization approximation. Our numerical examples in Section \ref{sec:ne} indicate that such implementation performs quite well for both low-dimensional and high-dimensional cases.

From the computational perspective, discretization approximation is non-iterative, and possesses a low computational complexity, requiring only $O(M)$ operations, which results in a remarkably short computing time.

\subsection{Sampling properties}\label{subsec:sample}

Sometimes we need random samples from the density $f$, which can also be generated from the discrete distribution $\mu_{(f,\s{A}_M)}$. For a sample size $N$, let $\v{X}_1,\ldots,\v{X}_N$ be independently and identically distributed according to \eqref{dp}. The sample $\{\v{X}_i\}_{i=1}^N$ can be viewed as a sample from $f$. Let $\mu_{(N,\s{A}_M)}$ denote its empirical probability measure. The following theorem indicates that $\mu_{(N,\s{A}_M)}$ is asymptotically equivalent to the empirical measure of the sample from $\mu_f$.

\begin{theorem}\label{th:rd} Suppose $\delta_{M,d}=O(1/\sqrt{N})$. Then \begin{equation}\label{d}d_\mathrm{K}(\mu_{(N,\s{A}_M)},\ \mu_f)=O_p(1/\sqrt{N}).\end{equation}
Suppose further $E_{(2k_1,\ldots,2k_d)}(\mu_f)<\infty$ and $V(f_{(k_1,\ldots,k_d)})<\infty$. Then \begin{equation}\label{de}\left|E_{(k_1,\ldots,k_d)}(\mu_{(N,\s{A}_M)})-E_{(k_1,\ldots,k_d)}(\mu_f)\right|=O_p(1/\sqrt{N}).\end{equation}\end{theorem}

Usually deterministic design strategies can improve the representativeness of $\mu_{(N,\s{A}_M)}$ over random sampling. Here we begin with the minimax distance design (Johnson, Moore, and Ylvisaker 1990) on $[0,1]$, $u_i=(2i-1)/(2N),\ i=1,\ldots,N$, and construct $\{\v{X}_i\}_{i=1}^N$ in the following deterministic manner. Let $q_0=0$ and $q_i=\sum_{j=1}^ip_j$ for $i=1,\ldots,M$. Clearly, $u_1<\cdots<u_N$ and $q_0\leqslant q_1\leqslant\cdots\leqslant q_M$.

For $j=1$, search from $\{1,\ldots,M\}$ in the increasing order, and find the minimal number $i_1$ such that $u_1\in[q_{i_1-1},q_{i_1})$, let $\v{X}_1=\v{a}_{i_1}$;

for $j=2$, search from $\{i_1,\ldots,M\}$ in the increasing order, and find the minimal number $i_2$ such that $u_2\in[q_{i_2-1},q_{i_2})$, let $\v{X}_2=\v{a}_{i_2}$;

$\vdots$

for $j=N$, search from $\{i_{N-1},\ldots,M\}$ in the increasing order, and find the minimal number $i_N$ such that $u_N\in[q_{i_N-1},q_{i_N})$, let $\v{X}_N=\v{a}_{i_N}$.

The above process actually provides a method to generate representation points (Mak and Joseph 2018) of the density $f$.
Let $\{\v{X}_i\}_{i=1}^N$ denote the set of these representation points. Then the corresponding empirical measure $\mu_{(N,\s{A}_M)}$ is non-stochastic. The following theorem shows its convergence properties.

\begin{theorem}\label{th:ewc} For sufficiently large $M$ and $N$, \begin{equation}\label{conv}d_\mathrm{K}(\mu_{(N,\s{A}_M)},\ \mu_f)\leqslant C\delta_{M,d}+C_0M/N,\end{equation} where $C$ and $C_0$ are positive constants independent of $N$, $M$, and $d$. Suppose further $E_{(2k_1,\ldots,2k_d)}(\mu_f)<\infty$ and $V(f_{(k_1,\ldots,k_d)})<\infty$. Then for sufficiently large $M$ and $N$, \begin{equation}\label{econv}\left|E_{(k_1,\ldots,k_d)}(\mu_{(N,\s{A}_M)})-E_{(k_1,\ldots,k_d)}(\mu_f)\right|\leqslant \tilde{C}\delta_{M,d}+\tilde{C}_0M/N,\end{equation}where $\tilde{C}$ and $\tilde{C}_0$ are positive constants independent of $N$, $M$, and $d$.\end{theorem}

By Theorem \ref{th:ewc}, we can obtain the deterministic convergence rate of the proposed representation points. As mentioned before, the quasi-Monte Carlo methods with higher-order sequences/digital nets give $\delta_{M,d}=C_d(\log{M})^{\alpha d}/M^\alpha$ if $f$ has smoothness of order $\alpha$. When $\alpha\geqslant2$, the optimal rate can be achieved by taking $N=c_dM^{\alpha+1}/(\log{M})^{\alpha d}$. Even with $M=N^{1/(\alpha+1)}$, we can get the rate $O((\log{N})^{\alpha d}/N^{\alpha/(\alpha+1)})$, which is deterministic and better than $O_p(1/\sqrt{N})$, the optimal statistical rate, achieved by the random sampling scheme in Theorem \ref{th:rd}.

The proposed method for constructing representation points requires $O(MN)$ operations, which is acceptable compared with optimization-based construction procedures (Mak and Joseph 2018). Note that there are repetitions in the representation points when $N>M$. For each point, we can add a random noise whose magnitude does not exceed the order $O(1/N)$, and this does not influence the convergence rate of these representation points.

\section{Approximation on general regions}\label{sec:unb}

When the support $\s{S}$ is unbounded, a popular strategy is to use a bounded set that contains the significant region of $f$ (Fu and Wang 2002; Joseph 2012). It is not easy to find such a set for many practical problems. In this section we propose a straightforward method to handle unbounded supports using integration transformation to the unit hypercube (Dick, Kuo, and Sloan 2013).

For a large integer $M$, take an initial point set $\s{A}_M^{(0)}=\{\v{a}_i\}_{i=1}^M\subset[0,1]^d$, say, a quasi-Monte Carlo sequence, as in the previous section.
Let $\psi(\v{x})$ be a density function (up to a constant) satisfying $\psi(\v{x})>0$ for $\v{x}\in\s{S}$. Suppose that there exists a one-to-one map $h$ defined on $[0,1]^d$ such that the random vector $\v{Y}=h(\v{X})\sim\psi(\v{x})$ for $\v{X}$ being uniformly distributed on $[0,1]^d$. By the integration equality, $$\int_{\mathbb{R}^d}f(\v{x})\mathrm{d}\v{x}=\int_{\mathbb{R}^d}\frac{f(\v{x})}{\psi(\v{x})}\psi(\v{x})\mathrm{d}\v{x}
=\int_{[0,1]^d}\frac{f\left(h(\v{x})\right)}{\psi\left(h(\v{x})\right)}\mathrm{d}\v{x},$$
we define a probability measure $\mu_{(f,\s{A}_M)}$ on the support point set $\s{A}_M=\{h(\v{a}_i)\}_{i=1}^M$ as
\begin{equation}\label{dpub}P\left(\v{X}_{(f,\s{A}_M)}=h(\v{a}_i)\right)=p_i
=\frac{f\left(h(\v{a}_i)\right)/\psi\left(h(\v{a}_i)\right)}{\sum_{j=1}^Mf\left(h(\v{a}_j)\right)/\psi\left(h(\v{a}_j)\right)},\ \ i=1,\ldots,M,\end{equation}
where the random vector $\v{X}_{(f,\s{A}_M)}\sim\mu_{(f,\s{A}_M)}$. As in Section \ref{sec:method}, $\mu_{(f,\s{A}_M)}$ is an approximation to $\mu_f$.
We can also construct the empirical measure $\mu_{(N,\s{A}_M)}$ through the random sampling and deterministic design strategies, and the corresponding convergence properties hold under regularity conditions. When $\s{S}=[0,1]^d$, \eqref{dpub} reduces to \eqref{dp} by taking $\psi$ as the uniform distribution on $[0,1]^d$.

Ideally we would like to choose a density function $\psi$, or, equivalently a transformation $h$, that leads to a good integrand in the unit hypercube. Similar to the proposal distribution in Monte Carlo methods such as the importance sampling and Metropolis-Hastings algorithm (Liu 2001), $\psi$ plays an important role in discretization approximation. Thus, we also call $\psi$ the proposal distribution/density. A feasible selection is the multivariate normal density with mean $\v{\mu}$ and positive definite covariance matrix $\m{\Sigma}$, which corresponds to
\begin{eqnarray*}&&\psi(\v{x})=\exp\left\{-\frac{1}{2}(\v{x}-\v{\mu})'\m{\Sigma}^{-1}(\v{x}-\v{\mu})\right\},\ \v{x}\in\mathbb{R}^d, \\&&h(\v{x})=\v{\mu}+\m{\Sigma}^{1/2}\left(\Phi^{-1}(x_1),\ldots,\Phi^{-1}(x_d)\right)',\ \v{x}=(x_1,\ldots,x_d)'\in[0,1]^d,\end{eqnarray*}
in \eqref{dpub}, where $\Phi(x)$ denotes the cumulative distribution function of the standard normal distribution. Actually, for a broader exploration, it may be better to take heavy-tailed density functions such as the multivariate Cauchy density, which gives
\begin{equation}\label{cau}
	\begin{aligned}
		&\psi(\v{x})=\phi_{\mathrm{cauchy}}\left(\m{\Sigma}^{-1/2}(\v{x}-\v{\mu})\right),\ \v{x}\in\mathbb{R}^d,  \\
		&h(\v{x})=\v{\mu}+\m{\Sigma}^{1/2}\left(\tan\left(\pi(x_1-1/2)\right),\ldots,\tan\left(\pi(x_d-1/2)\right)\right)',\ \v{x}=(x_1,\ldots,x_d)'\in[0,1]^d,
	\end{aligned}
\end{equation}where
$\phi_{\mathrm{cauchy}}(\v{x})=\phi_{\mathrm{cauchy}}(x_1,\ldots,x_d)=\prod_{i=1}^d(1+x_i^2)^{-1}$, $\v{\mu}$ denotes the vector of location parameters, and $\m{\Sigma}$ is a positive definite covariance matrix corresponding to the affine transformation. For variables defined on $(0,\infty)$ or a bounded $[L,U]$ in $f$, we can adopt a Gamma distribution or a uniform distribution on $[L,U]$ as the proposal distribution. For mixed variables, $\v{x}=(\v{x}^{(1)},\v{x}^{(2)},\v{x}^{(3)})$ with $\v{x}^{(1)}\in{\mathbb{R}}^{d_1},\ \v{x}^{(2)}\in(0,\infty )^{d_2}$, and $\v{x}^{(3)}\in\prod_{i=1}^{d_3}[L_i,U_i]$, we can take a product measure of multivariate Cauchy, Gamma, and uniform distributions.

In discretization approximation with \eqref{dpub}, we define the acceptance rate as \begin{equation}\label{r}R=\frac{\sum_{i=1}^MI(p_i>0)}{M},\end{equation}where $I$ denotes the indicator function. Similar to the acceptance rate in the Metropolis-Hastings algorithm, the acceptance rate in \eqref{r} is an important index to investigate the effectiveness of the proposal distribution $\psi$. A low rate, say, less than $0.1$, indicates that the significant region of the target distribution is not found, or that the exploration scope is too large. We can use a multi-stage adaptive procedure to implement discretization approximation. The first stage is exploration that searches the main region with an initial proposal distribution. After analyzing the support points and acceptance rate, the follow-up stages implement discretization approximation with modified proposal distributions. The fast speed of discretization approximation allows more than two stages for most cases.

To avoid numerical overflow, especially for high-dimensional problems, we usually calculate the negative log-density $\ell(\v{x})=-\log(f(\v{x}))$, instead of $f(\v{x})$ itself, and compute $f(\v{x})=\exp(\tau-\ell(\v{x}))$ in \eqref{dpub}, where $\tau$ is a parameter desired to be close to $\inf_{\v{x}\in\s{S}}\ell(\v{x})$. It can be seen that $\tau$ is related to the significant region of the target distribution. Therefore, the step of selecting $\tau$ can be included in the exploration stage.

\section{Numerical examples}\label{sec:ne}

\subsection{A simple one-dimensional example}\label{subsec:1d}
\begin{table}[t]\centering
\caption{Comparisons in Section \ref{subsec:1d} (standard deviations in parentheses)}
\begin{tabular}{lcccccc}
\hline
   &\quad& \multicolumn{2}{c}{$M=10$} &\quad& \multicolumn{2}{c}{$M=30$}
   \\\cline{3-4}\cline{6-7}   && (Mean) SE & (Mean) KD& & (Mean) SE & (Mean) KD \\ \hline
  MCMC           &&0.0121 (0.0156) &0.3304 (0.1090)& &0.0038 (0.0055) &0.1874 (0.0623) \\
  Exact MC       &&0.0082 (0.0115) &0.2700 (0.0814)& &0.0025 (0.0036) &0.1555 (0.0513) \\
  DA             &&2.1613$\times10^{-5}$     &0.0872         & &3.2417$\times10^{-7}$     &0.0275          \\
  DA-RP $(N=10)$ &&6.0494$\times10^{-5}$     &0.0951         & &1.2346$\times10^{-6}$     &0.0429          \\
 \hline
\end{tabular}\label{tb:betadis}
 \end{table}

In this example the target distribution is the mixture of two Beta distributions $0.5Beta(6,3)+0.5Beta(2,7)$. Three methods are compared: the Metropolis-Hastings algorithm in MCMC, the exact MC method, and discretization approximation (DA). In MCMC, the proposal distribution is the uniform distribution on $[0,1]$, which is consistent with that in the original DA in Section \ref{subsec:dc}. In DA, the support points are constructed from the minimax distance design on $[0,1]$. It should be pointed out that exact MC utilizes the mixture characteristic of the target distribution to generate random samples, while the other two methods only use the density expression.

For each method, squared error (SE) of the corresponding mean approximation and the Kolmogorov distance (KD) between the estimated distribution and the target distribution are computed. Table \ref{tb:betadis} reports the mean SEs and mean KDs over 100 repetitions of the two MC methods, and the SEs and KDs of DA with the same sizes $M=10$ and 30. It can be seen that DA has much better performance than the two MC methods. This is consistent with our theoretical results in Section \ref{subsec:dc}. We also construct the representation point (RP) set from DA by the deterministic procedure in Section \ref{subsec:sample}, and show its performance in Table \ref{tb:betadis}. With a small sample size $N=10$, the RP set also outperforms the MC methods.

\subsection{Two-dimensional examples}\label{subsec:2d}

\begin{table}[t]\centering\footnotesize
\caption{Comparisons of MSE (SE) in Section \ref{subsec:2d} (standard deviations in parentheses)}
\begin{tabular}{lcccccc}
\hline
   &\quad&$\v{\mu}$ & $\m{\Sigma}$& & $q_{0.2}$ of $X_1$& $q_{0.1}$ of $X_2$ \\ \hline
  MCMC ($M=2000$)              &&0.0018 (0.0018)      &0.0155 (0.0054)      & &0.0056 (0.0085)      &0.0029 (0.0037) \\
  Exact MC ($M=2000$)          &&0.0017 (0.0021)      &0.0259 (0.0395)      & &0.0050 (0.0074)      &0.0014 (0.0025) \\
  DA  ($M=1000$)               &&0.0021               &0.0040               & &5.4915$\times10^{-4}$ &0.0026          \\
  DA  ($M=2000$)               &&4.2451$\times10^{-5}$&5.5766$\times10^{-4}$& &4.4860$\times10^{-5}$&7.4497$\times10^{-6}$         \\
  two-stage DA ($M=1000+1000$) &&6.5859$\times10^{-7}$&3.0153$\times10^{-5}$& &2.0252$\times10^{-7}$&4.8824$\times10^{-6}$         \\
 \hline
\end{tabular}\label{tb:norm}
 \end{table}
\begin{figure}[t]\begin{center}
\scalebox{0.8}[0.8]{\includegraphics{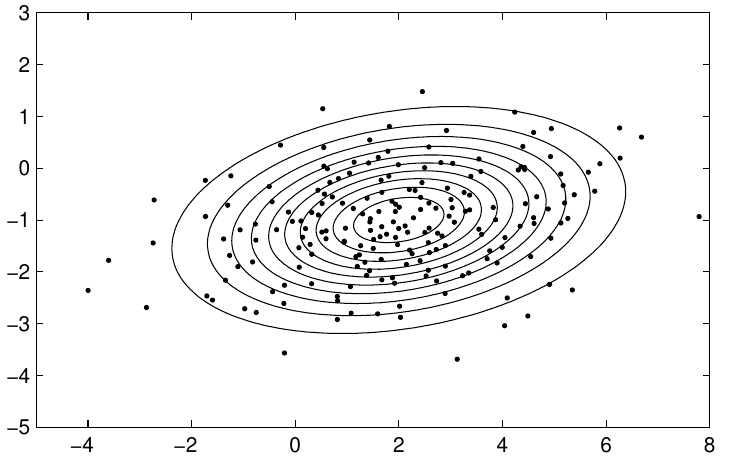}}
\end{center}
\caption{Contours of the two-dimensional normal density in Section \ref{subsec:2d} and the 200 representation points from two-stage DA. }\label{fig:norm}
\end{figure}

We consider the two-dimensional normal distribution with mean vector $\v{\mu}=(2,-1)'$ and covariance matrix $\m{\Sigma}=\begin{pmatrix} 4 & 1/2 \\ 1/2 & 1 \end{pmatrix}$, and also compare the three methods: the Gibbs sampling in MCMC, exact MC, and DA. The Gibbs sampling uses the conditional distributions to iteratively generate random variables. The sample sizes in the two MC methods are set as $M=2000$. In DA. we first use the two-dimensional Cauchy distribution with $\v{\mu}_{\mathrm{Cauchy}}=\v{0}$ and $\m{\Sigma}_{\mathrm{Cauchy}}=\m{I}_2$ in \eqref{cau} as the proposal distribution, and conduct DA with $M=1000$ and 2000. The support points are constructed by the Sobol' sequence (Dick, Kuo, and Sloan 2013) removing the first zero point. By replacing $\v{\mu}_{\mathrm{Cauchy}}$ and $\m{\Sigma}_{\mathrm{Cauchy}}$ with the mean and covariance estimates from this DA method with $M=1000$, we implement the second stage DA with $M=1000$.

For each method, SEs or mean squared errors (MSEs) over 100 repetitions of the corresponding approximations to $\v{\mu}$, $\m{\Sigma}$, and the 0.2-quantile or 0.1-quantile of the two marginal distributions are reported in Table \ref{tb:norm}. As expected, the DA methods have better performance than the two MC methods, even with a smaller size. Furthermore, the two-stage DA outperforms the original DA. We also construct the representation points from the two-stage DA by the deterministic procedure in Section \ref{subsec:sample}, and show them in Figure \ref{fig:norm}.

\begin{figure}[t]\begin{center}
\scalebox{0.7}[0.7]{\includegraphics{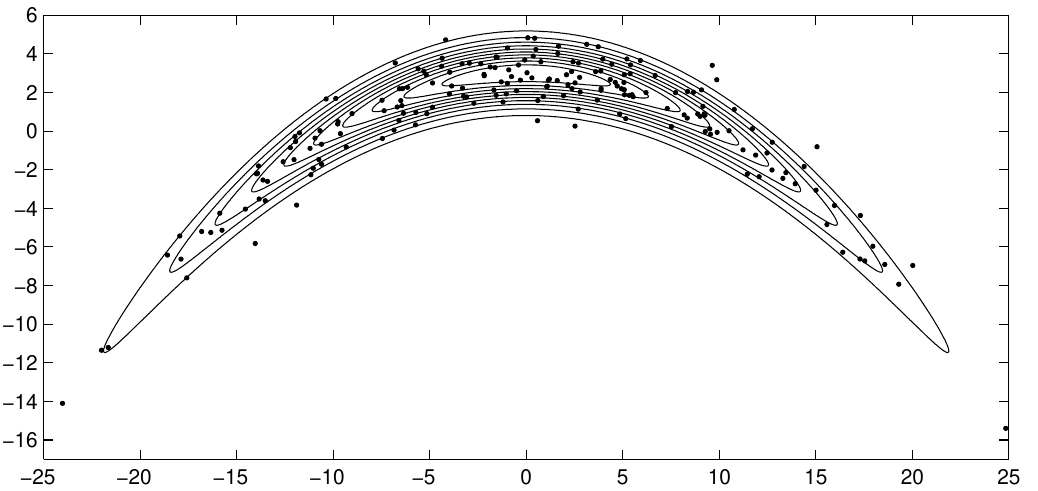}}
\end{center}
\caption{Contours of the banana density in Section \ref{subsec:2d} and the 200 representation points from two-stage DA. }\label{fig:banana}
\end{figure}

We next consider the two-dimensional density with banana-shaped contours, $f(x_1,x_2)=\phi(x_1;\ 0, 100)\phi(x_2+0.03x_1^2-3;\ 0,1)$, where $\phi(\cdot;\ \mu,\sigma^2)$ denotes the density function of the one-dimensional normal distribution with mean $\mu$ and variance $\sigma^2$. This density is commonly used in evaluation of Bayesian algorithms (Haario, Saksman, and Tamminen 2001; Joseph 2012). To approximate its mean vector, the proposed two-stage DA with $M=10^6+10^6$ gives SE $1.0376\times10^{-4}$, and the computing time is about 0.4 seconds, while MCMC and the Gaussian process interpolation method (Joseph 2012) take more than one minutes to achieve such accuracy. The 200 representation points generated by  the deterministic procedure in Section \ref{subsec:sample} are depicted in Figure \ref{fig:banana}.

\subsection{The linear regression}\label{subsec:lm}

Consider the linear regression model
\begin{equation}y_i=\beta_0+\beta_1x_{i1}+\cdots+\beta_px_{id}+\varepsilon_i=\beta_0+{\mathbf{x}}_i'\v{\beta}+\varepsilon_i,\
i=1,\ldots,n,\label{lm}\end{equation} where $\{(\v{x}_i,y_i)\}_{i=1}^n$ are observations, $\v{x}_i=(x_{i1},\ldots,x_{id})'$, $\beta_0$ and $\v{\beta}=(\beta_1,\ldots,\beta_d)'$ are unknown coefficients, and $\varepsilon_1,\ldots,\varepsilon_n$ are independently and identically distributed normal random errors with mean 0 and variance $\sigma^2$. Denote $\m{X}=(x_{ij})_{i=1,\ldots,n,\ j=1,\ldots,d},\ \m{Z}=(\v{1}_n\ \m{X}),\ \v{1}_n=(1,\ldots,1)',\ \v{y}=(y_1,\ldots,y_n)'$, and $\v{\gamma}=(\beta_0,\v{\beta}')'$. We get the likelihood \begin{equation*}\v{y}\mid\v{\gamma},\sigma^2\sim N(\m{Z}\v{\gamma},\sigma^2\m{I}_n).\end{equation*}Take the non-informative prior (Gelman et al. 2004) $$p(\v{\gamma},\sigma^2)\propto \frac{1}{\sigma^2}I\left(\sigma^2>0\right).$$We have the posterior density \begin{equation}\label{lmpos}p(\v{\gamma},\sigma^2\mid \v{y})\propto(\sigma^2)^{-n/2-1}\exp\left[-\frac{1}{2\sigma^2}\|\v{y}-\m{Z}\v{\gamma}\|^2\right]I\left(\sigma^2>0\right),\end{equation}where $\hat{\v{\gamma}}=(\m{Z}'\m{Z})^{-1}\m{Z}'\v{y}=(\hat{\beta}_0,\hat{\beta}_1,\ldots,\hat{\beta}_d)'$ and $s^2=(\v{y}-\m{Z}\hat{\v{\gamma}})'(\v{y}-\m{Z}\hat{\v{\gamma}})/(n-d-1)$ under the assumption $\m{Z}$ is of full column rank. Let $\mathrm{Inv}\chi^2_{m}$ denote the inverse $\chi^2$ distribution of degree of freedom $m$. Some algebra yields
\begin{equation}\label{lml}\sigma^2\mid\v{y}\sim (n-d-1)s^2\mathrm{Inv}\chi^2_{n-d-1},\quad \v{\gamma}\mid \v{y},\sigma^2\sim N(\hat{\v{\gamma}},\ \sigma^2(\m{Z}'\m{Z})^{-1}),\end{equation}which can be used to compute the posterior means and quantiles of the parameters.
\begin{table}[t]\centering
\caption{Comparisons of MSE in Section \ref{subsec:lm} (standard deviations in parentheses)}
\begin{tabular}{lcccccc}
\multicolumn{7}{c}{$d=100$}\\
\hline
   &\quad& \multicolumn{2}{c}{posterior of $\beta_1$} &\quad& \multicolumn{2}{c}{posterior of $\sigma^2$}
   \\\cline{3-4}\cline{6-7}   && mean ($\times 10^{-2}$) &$q_{0.1}$ ($\times 10^{-1}$)  & &mean ($\times 10^{-3}$) &$q_{0.1}\ (\times 10^{-2})$  \\ \hline
  MCMC           &&7.7625 (14.158) &1.1151 (1.9518)& &9.4970 (16.504) &4.1879 (4.5843) \\
  %Exact MC       &&0.0005 (0.0006) &0.0844 (0.0272)& &0.0232 (0.0342) &0.0042 (0.0064) \\
  DA             &&0.1141 (0.1027) &0.0180 (0.0117)& &0.1476 (0.0422) &0.7232 (0.2346) \\
 \hline\\[-2mm]\multicolumn{7}{c}{$d=1000$}\\
\hline
   &\quad& \multicolumn{2}{c}{posterior of $\beta_1$} &\quad& \multicolumn{2}{c}{posterior of $\sigma^2$}
   \\\cline{3-4}\cline{6-7}   && mean ($\times 10^{-7}$) &$q_{0.1}$ ($\times 10^{-2}$)  & &mean ($\times 10^{-4}$) &$q_{0.1}\ (\times 10^{-4})$  \\ \hline
  MCMC           &&365.36 (531.90) &2.9714 (0.6836)& &282.12 (10.657) &5.6816 (11.318) \\
  %Exact MC       &&0.9512 (1.4938) &2.1010 (0.4470)& &0.0219 (0.0314) &0.0370 (0.0432) \\
  DA             &&0.0001 (0.0001) &2.9607 (0.5975)& &4.3145 (1.3108) &5.6525 (10.997) \\
 \hline
\end{tabular}\label{tb:beta}
 \end{table}

We conduct a simulation study to compare DA and MCMC under \eqref{lm}. Set $\beta_0=3$, $\v{\beta}=(-1,\ 2,\ 1.5,\ 0.\ldots,0)'$, and $\sigma^2=1$. The predictors $\v{x}_1,\ldots,\v{x}_n$ are independently generated from $N\left(\v{0},\ 0.5\v{1}_d\v{1}_d'+0.5\m{I}_d\right)$. We consider two values of $d$, 100 and 1000, and let $n=d+100$. The Metropolis-Hastings algorithm and DA both use the density expression \eqref{lmpos} only. The support points in DA are constructed from the Sobol' sequence.

For $d=100$, the proposal distributions in the Metropolis-Hastings algorithm and DA are both taken as the product of the multivariate Cauchy distribution with $\v{\mu}_{\mathrm{Cauchy}}=\hat{\v{\gamma}}$ and $\m{\Sigma}_{\mathrm{Cauchy}}=s^2(\m{Z}'\m{Z})^{-1}$ for $\v{\gamma}$ and the Gamma distribution with shape parameter $n-d-1$ and scale parameter $s^2/(n-d-1)$ for $\sigma^2$. True values of the posterior distributions of $\beta_1$ and $\sigma^2$ can be exactly computed by \eqref{lml}. We use the two methods with $M=1000$ to approximate the means and $0.1$-quantiles of their posterior distributions, and report the MSEs over 1000 repetitions in Table \ref{tb:beta}. It can be seen that. with the same proposal distribution, DA is much better than MCMC.

For $d=1000$, the above proposal distribution leads to very low acceptance rates for both MCMC and DA. We try several different proposal distributions in DA, and select the uniform distribution on $[\hat{\beta}_0-0.02,\ \hat{\beta}_0+0.02]\times[\hat{\beta}_1-0.02,\ \hat{\beta}_1+0.02]\times\cdots\times[\hat{\beta}_d-0.02,\ \hat{\beta}_d+0.02]\times[s^2-0.15,\ s^2+0.15]$. Due to the fast speed of DA, the computing time of this process is no more than ten seconds. We also use this proposal distribution in MCMC. The simulation results with $M=10000$ are shown in Table \ref{tb:beta}. It can be observed that such a proposal distribution also works for MCMC. Compared with MCMC, DA has a much better performance in mean approximation and slightly better performance in quantile approximation.

\begin{table}[t]\centering
\caption{CR and AL of $90\%$ credible intervals in Section \ref{subsec:lm} (standard deviations in parentheses)}
\begin{tabular}{lcccccc}
\hline
   &\quad& \multicolumn{2}{c}{$\beta_1$} &\quad& \multicolumn{2}{c}{$\beta_4$}
   \\\cline{3-4}\cline{6-7}   && CR &AL& &CR&AL\\ \hline
  $n=50,\ d=5$          &&0.8524 &0.6141 (0.0924)& &0.9080 &1.4479 (0.3200) \\
  $n=100,\ d=10$        &&0.8848 &0.4476 (0.0497)& &0.5367 &1.1036 (0.2182) \\
  $n=200,\ d=20$        &&0.3881 &0.1385 (0.0786)& &0.9753 &1.3267 (0.1489) \\
 \hline
\end{tabular}\label{tb:cr}
 \end{table}

It is known that many Bayesian methods have good frequentist properties, especially in large-sample cases (Ghosh, Delampady, and Samanta 2006). However, for complex problems, computational difficulties hamper sufficient investigation of Bayesian methods' finite-sample frequentist properties that require a large number of repeated numerical simulations. This requirement may be met with discretization approximation because of its fast speed. Here we conduct simulations to study frequentist properties of the Bayesian lasso method (Park and Casella 2008) under \eqref{lm}. The lasso estimate (Tibshirani 1996) is the solution to minimize $$\|\v{y}-\beta_0\v{1}_n-\m{X}\v{\beta}\|^2+\lambda\sum_{j=1}^d|\beta_j|,$$where $\lambda>0$ is a tuning parameter. It can also be viewed as the Bayesian posterior mode by assigning the Laplace prior for $\v{\beta}$,$$p(\v{\beta}\mid \lambda)\propto \lambda\exp\left(-\lambda\sum_{j=1}^d|\beta_j|\right).$$For $\lambda$ and other parameters, we adopt a hierarchical structure of the priors$$p(\beta_0,\sigma^2)\propto \frac{1}{\sigma^2}I\left(\sigma^2>0\right),\quad\lambda\mid\sigma^2\sim U\left[0,\ 4\sigma\sqrt{n\log(d)}\right].$$ The selection of $\lambda$'s prior follows Wainwright (2019). Then the joint posterior is\begin{eqnarray*}p(\v{\beta},\beta_0,\sigma^2,\lambda\mid \v{y})&\propto&(\sigma^2)^{-n/2-1}\exp\left(-\frac{1}{2\sigma^2}\|\v{y}-\beta_0\v{1}_n-\m{X}\v{\beta}\|^2-\lambda\sum_{j=1}^d|\beta_j|\right)
\\&&\cdot I\left(\lambda\in\left[0,\ 4\sigma\sqrt{n\log(d)}\right],\ \sigma^2>0\right)\end{eqnarray*}

Under the same settings in Table \ref{tb:beta}, we use discretization approximation with $M=10000$ to compute $90\%$ credible intervals of $\beta_1$ (zero) and $\beta_4$ (nonzero), and show their coverage rates (CRs) and average lengths (ALs) over 10000 repetitions in Table \ref{tb:cr}. It can be seen that the Bayesian intervals perform well only for small $d$. These computing results with the fast Bayesian algorithm may indicate that a unified $\lambda$ cannot guarantee frequentist properties. The Bayesian lasso method with adaptive priors of $\lambda$ (Zou 2006) and the corresponding theoretical investigation seem valuable topics in the future.

\subsection{The full Bayesian Gaussian process regression}\label{subsec:gp}
The Gaussian process model, also called the Kriging model, is widely used in spatial statistics (Cressie 2015), computer experiments (Santner, Williams and Notz 2018), and machine learning (Rasmussen and Williams 2006). We consider the following Gaussian process regression model
\begin{equation}\label{GPmodel}
	y_i=\v{g}(\v{x}_i)'\v{\beta}+Z(\v{x}_i)+\varepsilon_i,\ i=1,\ldots,n,
\end{equation}
where $\v{g}(\v{x})=(g_{1}(\v{x}),\ldots,g_{q}(\v{x}))'$ is a set of pre-specified basis functions, $\v{\beta}=(\beta_1,\ldots,\beta_q)'$ is a vector of unknown regression coefficients, $\varepsilon_1,\ldots,\varepsilon_n$ are independently and identically distributed normal random errors with mean 0 and variance $\sigma_\varepsilon^2$, and $Z(\cdot)$ is a Gaussian process on a $d$-dimensional input domain $\s{X}$, denoted by $Z(\cdot)\sim \mathrm{GP}(\v{0},\ \sigma^2R(\cdot))$,
with variance $\sigma^{2}$ and correlation function $R(\cdot)$. Here we choose the Gaussian correlation
\begin{equation}\label{gasp_corr}
	R(\v{x}_{1}-\v{x}_{2})=R(\v{x}_{1}-\v{x}_{2}\mid\v{\eta})=\exp\left[-\sum_{i=1}^{d}\eta_{i}({x}_{1i}-{x}_{2i})^2\right],
\end{equation}
for $\v{x}_1=(x_{11},\ldots,x_{1d})',\ \v{x}_2=(x_{21},\ldots,x_{2d})'$, where $\v{\eta}=(\eta_1,\ldots,\eta_d)'$ is the vector of correlation parameters with $\eta_i>0$ for $i=1,\ldots,d$. Let $\rho=\sigma^2_{\varepsilon}/\sigma^2$ be the nugget-variance ratio, and $\v{\theta}=(\v{\beta}',\sigma^2,\v{\eta}',\rho)'$.
Therefore, the vector of response values $\v{y}=(y_1,\ldots,y_{n})'$ follows a multivariate normal distribution
\begin{equation*}
	\v{y}\mid\v{\theta}\sim N\left(\m{G}\v{\beta},\sigma^2\m{\Sigma}\right),
\end{equation*}
where $\m{G}=[\v{g}(\v{x}_1),\ldots,\v{g}(\v{x}_n)]'$,  $\m{\Sigma}=\m{R}+\rho\m{I}_n$ denotes the correlation matrix, $\m{R}$ is the matrix with the $(i,j)$th element $R(\v{x}_{i}-\v{x}_{j})$ defined by \eqref{gasp_corr}.

In most Bayesian analyses under Gaussian process models, the parameters $\v{\eta}$ and $\rho$ in the correlation matrix are treated as fixed values and usually replaced by their estimates. This strategy often underestimates the true uncertainty (Zimmerman and Cressie 1992).
Wu, Xiong, and Chien (2026) proposed a full Bayesian method for \eqref{GPmodel} that incorporates prior information for the correlation parameters. Their joint prior is
\begin{equation*}
	p(\v{\theta})\propto \frac{1}{\sigma^2}\exp\left[-\frac{1}{2}\left(\sum_{j=1}^d\eta_j+\rho\right)\right]I\left(\sigma^2>0,\ \rho>0,\ \eta_j>0,\ j=1,\ldots,d\right),
\end{equation*}which leads to the posterior
\begin{eqnarray}\label{thetapos}
p(\v{\theta}\mid\v{y})&\propto&(\sigma^2)^{-n/2-1}|\m{\Sigma}|^{-1/2}I\left(\sigma^2>0,\ \rho>0,\ \eta_j>0,\ j=1,\ldots,d\right)\nonumber
\\&& \cdot\exp\left[-\frac{1}{2\sigma^2}(\v{y}-\m{G}\v{\beta})'\m{\Sigma}^{-1}(\v{y}-\m{G}\v{\beta})-\frac{1}{2}\left(\sum_{j=1}^d\eta_j+\rho\right)\right].
\end{eqnarray}Our discretization approximation can be used to compute the posterior distributions of the parameters based on \eqref{thetapos}.

 In many applications, one needs to compute the prediction distribution of the response value $y^*$ at a new input value $\v{x}^*$.
Let $\v{g}_*=\v{g}(\v{x}^*)$ and $\v{r}_{*}=(R(\v{x}^*-\v{x}_1),\ldots,R(\v{x}^*-\v{x}_n))'$. Under \eqref{GPmodel}, we have
\begin{equation*}
	y^*\mid\v{y},\v{\theta} \sim N\left(\mu(\v{x}^*,\v{\theta}),\omega(\v{x}^*,\v{\theta})\right),
\end{equation*}
where \begin{equation}\label{musig}
	\mu(\v{x}^*,\v{\theta})=\v{g}_*'\v{\beta}+\v{r}_{*}'\m{\Sigma}^{-1}(\v{y}-\m{G}\v{\beta}),\quad
	\omega(\v{x}^*,\v{\theta})=\sigma^2(1+\rho-\v{r}_{*}'\m{\Sigma}^{-1}\v{r}_{*}).
\end{equation}
Then the posterior predictive density is \begin{equation}\label{ppd}
	p(y^*\mid\v{y})=\int p(y^*\mid\v{y},\v{\theta})\,p(\v{\theta}\mid\v{y})\mathrm{d}\v{\theta}.
\end{equation}With discretization approximation, \eqref{ppd} can be approximated by $$\hat{p}(y^*\mid\v{y})=\sum_{i=1}^Mp_i\phi(y^*;\ \mu(\v{x}^*,\v{\theta}_i),\omega(\v{x}^*,\v{\theta}_i)),$$where $\{\v{\theta}_i\}_{i=1}^M$ is the support point set generated by \eqref{dpub} and $p_i$ is the corresponding probability at $\v{\theta}_i$. By taking expectation, we can get the point prediction
\begin{equation}\label{pre}\hat{y}^*=\sum_{i=1}^Mp_i\mu(\v{x}^*,\v{\theta}_i).\end{equation}

The cost of computing the inverse of $\m{\Sigma}$ in \eqref{thetapos} is unacceptable for large $n$, especially in time-consuming Bayesian computation. The random Fourier features method (Rahimi and Recht 2007) is feasible to alleviate this problem, and can be described briefly as follows.
Draw $d$-dimensional random vectors $\v{\omega}_1,\ldots,\v{\omega}_m$ independently and identically distributed according to $N(\v{0}, \m{\Sigma}_{\v{\eta}})$, where $\m{\Sigma}_{\v{\eta}}=\mathrm{diag}\left\lbrace 2\eta_1,\cdots,2\eta_d\right\rbrace$.
Let $\m{\Pi}=(\v{\omega}_1,\ldots,\v{\omega}_m)$ and $\m{Z}_m=(\cos(\m{X\Pi}),\;\sin(\m{X\Pi}))/\sqrt{m}$. Thus $\m{\Sigma}$ is approximated by $\hat{\m{\Sigma}}=\m{Z}_m\m{Z}_m'+\rho\m{I}_n$, and $\m{\Sigma}^{-1}$ in \eqref{thetapos} and \eqref{musig} can be replaced by \begin{equation}\label{rff}(\m{Z}_m\m{Z}_m'+\rho\m{I}_n)^{-1}=\rho^{-1}\left[ \m{I}_{n}-\m{Z}_m\left(\m{Z}_m'\m{Z}_m+\rho\m{I}_{2m}\right)^{-1}\m{Z}_m'\right].\end{equation} Wu, Xiong, and Chien (2026) developed a random Fourier features-based MCMC algorithm to compute the posterior \eqref{thetapos}. A real dataset from a metro simulation (Li, Cui, and Xiong 2023) was used to illustrate the algorithm.

\begin{table}[t]\centering
\caption{Comparisons on the metro simulation dataset in Section \ref{subsec:gp}}
\begin{tabular}{lcccc}
\hline
   &\quad&squared prediction error & \quad&computing time (h.)\\ \hline
  MCMC  &&0.7820&&43.06 \\
  VI    &&1.1225&&6.208 \\
  DA    &&0.7744&&4.182 \\
 \hline
\end{tabular}\label{tb:gp}
 \end{table}

Here we use discretization approximation to conduct the prediction by \eqref{pre} with the random Fourier features simplification, and apply it to the above metro simulation dataset. The dataset uses six input variables to predict a response, the travel time of each passenger. It contains 8801 independent runs, with $n=5000$ used as the training set and the remaining $3801$ runs as the test set. As in Wu, Xiong, and Chien (2026), take $\v{g}(\v{x})=(1,\ \v{x}')'$ in \eqref{GPmodel} and $m=150$ in \eqref{rff} in our method. The number of support points $M$ is set as 2000. In Table \ref{tb:gp}, the corresponding squared prediction error and computing time are compared with the MCMC and variational inference (VI) methods in Wu, Xiong, and Chien (2026). Even with the random Fourier features simplification, it is time-consuming to calculate a value of the density in \eqref{thetapos} that involves matrix computations. This influences the computation speeds of the three methods. It can be seen that the proposed algorithm takes the shortest time.

\section{Discussion}\label{sec:dis}

In this paper we have proposed the discretization approximation approach for Bayesian computation when only the density function is available. Discretization approximation has a solid theoretical foundation from numerical integration theory. It can possess fast convergence rate for fixed $d$, or a convergence rate independent of $d$. In addition, discretization approximation is deterministic, non-iterative, flexible, and easy to implement with fast convergence and low computational cost. Numerical comparisons indicate that it outperforms MCMC in terms of both accuracy and speed for many problems. \texttt{MATLAB} codes for the numerical examples are available online or from the author.

In our opinion, the greatest advantage of discretization approximation is simplicity: it can be understood and used by those who know elementary statistics only. This point implies its potential for future broad applications. Besides,
the features of discretization approximation in accuracy and speed can make it possible to run a large number of simulations for Bayesian methods, which is helpful to study their frequentist properties; see Table \ref{tb:cr} in Section \ref{subsec:lm}. This is a point of considerable interest to many practitioners.

It is known that quasi-Monte Carlo and other space-filling designs are commonly used for traditional experimental design purposes such as numerical integration and input control, This work shows that they can also be used for distribution approximation. The theoretical findings in Section \ref{sec:method} indicate that the error bound of discretization approximation is actually that of numerical integration of the support points, i.e., the discrepancy. Note that the discrepancy is a space-filling criterion in experimental design, and that there are other designs with lower discrepancy than quasi-Monte Carlo sequences we use in this paper (Fang, Li, and Sudjianto 2006). For high-dimensional integration, another competing technique is based on sparse grid methods (Bungartz and Griebel 2004). Therefore, selection of support points in discretization approximation for specific problems is still a valuable topic in the future.

MCMC, as an indispensable tool in Bayesian statistics, has been widely studied, extended, and applied. Today it is still an active research direction. In fact, MC methods can be viewed as special cases of discretization approximation since they use the empirical distribution, which is discrete, of the generated sample to compute the posterior quantities. The author is not an expert in this area. We expect that discretization approximation can be a good complement to MCMC. Indeed, our numerical simulations in Section \ref{subsec:lm} show that they share the same proposal distributions, and that discretization approximation can help MCMC find a suitable one. In the future, discretization approximation can be combined with MCMC or other methods such as variational inference to develop more effective algorithms in Bayesian computation. We can also apply the discretization idea to other problems (Yi, Liu, Liu, et al. 2023; Zhao, Liu, Du, et al. 2025).

\section*{Appendix}

\emph{Proof of Theorem \ref{th:wc}}\quad First consider \eqref{feb}. Denote $J=\int_{[0,1]^d}f(\v{x})\mathrm{d}\v{x}\in(0,\infty)$. Let $I$ represent the indicator function. For any $\v{x}\in[0,1]^d$, let $J_{\v{x}}=\int_{[\v{0},\v{x}]}f(\v{z})\mathrm{d}\v{z}$. We have
\begin{eqnarray*}&&\mu_{(f,\s{A}_M)}\left([\v{0},\v{x}]\right)=\sum_{i=1}^Mp_iI\left(\v{a}_i\in[\v{0},\v{x}]\right)\\&=&\left(\frac{1}{M}\sum_{i=1}^Mf(\v{a}_i)\right)^{-1}
\cdot\frac{1}{M}\sum_{i=1}^Mf(\v{a}_i)I\left(\v{a}_i\in[\v{0},\v{x}]\right)
\\&=&(J+\Delta_1)^{-1}(J_{\v{x}}+\Delta_2),\end{eqnarray*} where \begin{equation}\label{delta}\Delta_1=\frac{1}{M}\sum_{i=1}^Mf(\v{a}_i)-J,\quad \Delta_2=\frac{1}{M}\sum_{i=1}^Mf(\v{a}_i)I\left(\v{a}_i\in[\v{0},\v{x}]\right)-J_{\v{x}}.\end{equation}By \eqref{eb}, $|\Delta_1|,\ |\Delta_2|\leqslant\delta_{M,d}V(f)$. Therefore, for sufficiently large $M$, \begin{eqnarray*}&& \left|\mu_{(f,\s{A}_M)}\left([\v{0},\v{x}]\right)-\mu_f\left([\v{0},\v{x}]\right)\right|
=\left|(J+\Delta_1)^{-1}(J_{\v{x}}+\Delta_2)-J^{-1}J_{\v{x}}\right|\\&=&\frac{J|\Delta_1|+J_{\v{x}}|\Delta_2|}{J|J+\Delta_1|}
\leqslant\frac{J|\Delta_1|+J_{\v{x}}|\Delta_2|}{J^2/2}\leqslant\frac{2V(f)}{J}\delta_{M,d},\end{eqnarray*}which implies \eqref{feb}.

Denote $J_{\v{k}}=\int_{[{0},1]^d}x_1^{k_1}\cdots x_d^{k_d}f(\v{x})\mathrm{d}\v{x}<\infty$.
We have
$$E_{\v{k}}(\mu_{(f,\s{A}_M)})=\left(\frac{1}{M}\sum_{i=1}^Mf(\v{a}_i)\right)^{-1}\cdot\frac{1}{M}\sum_{i=1}^Ma_{i1}^{k_1}\cdots a_{id}^{k_d}f(\v{a}_i)
=(J+{\Delta}_1)^{-1}(J_{\v{k}}+\tilde{\Delta}_2),$$ where $\v{a}_i=(a_{i1},\ldots,a_{id})'$ for $i=1,\ldots,M$, $\Delta_1$ is defined in \eqref{delta}, and \\$\tilde{\Delta}_2=\sum_{i=1}^Ma_{i1}^{k_1}\cdots a_{id}^{k_d}f(\v{a}_i)/M-J_{\v{k}}$. By \eqref{eb}, the bounds of $\Delta_1$ and $\tilde{\Delta}_2$
imply \eqref{jfeb}.  \qed

\emph{Proof of Theorem \ref{th:rd}}\quad For \eqref{d}, it suffices to prove $d_\mathrm{K}(\mu_{(N,\s{A}_M)},\ \mu_{(f,\s{A}_M)})=O_p(1/\sqrt{N})$. By DKW's inequality, for any $\epsilon, z>0$ and $N$, there exists a positive constant $C_{\epsilon,d}$, not depending on $M$, such that $P\left(d_\mathrm{K}(\mu_{(N,\s{A}_M)},\ \mu_{(f,\s{A}_M)})>z\right)\leqslant C_{\epsilon,d}e^{-2(2-\epsilon)Nz^2}$, which implies $E\left[\sqrt{N}d_\mathrm{K}(\mu_{(N,\s{A}_M)},\ \mu_{(f,\s{A}_M)})\right]^2=O(1)$ (Shao 1999). This completes the proof of \eqref{d}.

Next consider \eqref{de}. It suffices to prove $\left|E_{(k_1,\ldots,k_d)}(\mu_{(N,\s{A}_M)})-E_{(k_1,\ldots,k_d)}(\mu_{(f,\s{A}_M)})\right|=O_p(1/\sqrt{N})$. By Bernstein's inequality (Massart 2006), for any $z>0$ and $N$,
\\$P\left(\left|E_{(k_1,\ldots,k_d)}(\mu_{(N,\s{A}_M)})-E_{(k_1,\ldots,k_d)}(\mu_{(f,\s{A}_M)})\right|>z\right)\leqslant\exp\left\{-Nz^2/(2\sigma_M^2+4z/3)\right\}$,
where $\sigma_M^2=E_{(2k_1,\ldots,2k_d)}(\mu_{(f,\s{A}_M)})-\left[E_{(k_1,\ldots,k_d)}(\mu_{(f,\s{A}_M)})\right]^2\leqslant E_{(2k_1,\ldots,2k_d)}(\mu_{(f,\s{A}_M)})
\leqslant E_{(2k_1,\ldots,2k_d)}(\mu_f)+1$ for sufficiently large $M$. Similar to the proof of \eqref{d}, we can get \\$E\left[\sqrt{N}\left|E_{(k_1,\ldots,k_d)}(\mu_{(N,\s{A}_M)})-E_{(k_1,\ldots,k_d)}(\mu_{(f,\s{A}_M)})\right|\right]^2=O(1)$, which completes the proof.\qed

\emph{Proof of Theorem \ref{th:ewc}}\quad
We first prove that, for any $\v{a}_i\in\s{A}_M$, \begin{equation}\label{pc} \left|\mu_{(N,\s{A}_M)}\left(\{\v{a}_{i}\}\right)-\mu_{(f,\s{A}_M)}\left(\{\v{a}_{i}\}\right)\right|
=\left|\mu_{(N,\s{A}_M)}\left(\{\v{a}_{i}\}\right)-p_i\right|\leqslant C_0/N.\end{equation}
This holds obviously if $p_i=0$. We next consider $p_i>0$. First consider $i=1$. Take $\tilde{i}=[Np_1+1/2]$, where $[\cdot]$ represents the floor function, and we have $(2\tilde{i}-1)/(2N)<p_1$ and $(2\tilde{i}+1)/(2N)\geqslant p_1$, which indicates that there are $\tilde{i}$ repeated $\v{a}_1$'s in $\{\v{X}_i\}_{i=1}^N$. Therefore, $\mu_{(N,\s{A}_M)}\left(\{\v{a}_1\}\right)=\tilde{i}/N=[Np_1+1/2]/N$, and $\left|\mu_{(N,\s{A}_M)}\left(\{\v{a}_{1}\}\right)-p_1\right|\leqslant 3/(2N)$, which implies \eqref{pc}. The case of $i>1$ can be similarly discussed. By \eqref{pc}, $d_\mathrm{K}(\mu_{(N,\s{A}_M)},\ \mu_{(f,\s{A}_M)})\leqslant C_0M/N$. Therefore. we get \eqref{conv} by combining this and Theorem \ref{th:wc}.

Next consider \eqref{econv}. Suppose that there are $\tilde{i}_j$ repeated $\v{a}_j$'s in $\{\v{X}_i\}_{i=1}^N$ for $j=1,\ldots,M$. Similar to the above proof, $|\tilde{i}_j/N-p_j|\leqslant \tilde{C}_0/N$ for all $j$, where $\tilde{C}_0$ is a positive constant independent of $N$, $M$, and $d$. We have
$\left|E_{(k_1,\ldots,k_d)}(\mu_{(N,\s{A}_M)})-E_{(k_1,\ldots,k_d)}(\mu_{(f,\s{A}_M)})\right|=\left|\sum_{j=1}^M\tilde{i}_ja_{j1}^{k_1}\cdots a_{jd}^{k_d}/N-\sum_{j=1}^Mp_ja_{j1}^{k_1}\cdots a_{jd}^{k_d}\right|
\leqslant\sum_{j=1}^M\left|\tilde{i}_j/N-p_j\right| a_{j1}^{k_1}\cdots a_{jd}^{k_d}\leqslant\tilde{C}_0M/N$, which completes the proof by Theorem \ref{th:wc}.
\qed

\section*{Supplementary Materials}
\texttt{codes.zip}: contains the \texttt{MATLAB} codes for the numerical examples in Section \ref{sec:ne}.

\section*{Data Availability Statement}
Most data supporting the findings of this study were generated by statistical simulations.
Raw data in Section \ref{subsec:gp} were generated by a metro simulation. Derived data are available from the Supplementary Materials.

\section*{Disclosure Statement}
No potential conflict of interest was reported by the author.

\section*{Acknowledgments}
This work is partially supported by Strategic Priority Research Program of Chinese Academy of Sciences (grant number XDA0500200) and National Natural Science Foundation of China (grant number 12571276).

\vspace{1cm} \noindent{\Large\bf References}

{\begin{description}
\footnotesize

\item
Blei, D. M., Kucukelbir, A., and McAuliffe, J. D. (2017). Variational Inference: A Review for Statisticians. Journal of the American Statistical Association 112, 859-877.

\item
Bornkamp, B. (2011), Approximating probability densities by iterated Laplace approximations, Journal of Computational and Graphical Statistics,
20, 656--669.

Bungartz, H. and Griebel, M. (2004), Sparse grids. \textit{Acta Numerica}, 13, 147--269.

\item
Cipra, B.A. (2000) The best of the 20th century: editors name top 10 algorithms. SIAM News 33(4), 1-2.

\item
Conrad, P. R., Marzouk, Y. M., Pillai, N. S., and Smith, A. (2016), Accelerating asymptotically exact MCMC for computationally intensive models via local approximations, Journal of the American Statistical Association, 111:516, 1591-1607

\item{}
Cressie, N. A. C. (2015). \textit{Statistics for Spatial Data}, Revised Edition. John Wiley \& Sons.

\item
Dick, J. (2009). On quasi-Monte Carlo rules achieving higher order convergence. In Monte Carlo and Quasi-Monte Carlo Methods 2008 (pp. 73-96). Berlin, Heidelberg: Springer Berlin Heidelberg.

\item
Dick, J., Kuo, F. Y., and Sloan, I. H. (2013), High-dimensional integration: The quasi-Monte Carlo way. \textit{Acta Numerica}, 22, 133--288.

%\item
%Fang, K. T. and Wang, Y. (1993), \textit{Number-theoretic methods in statistics}. CRC Press.

\item
Fang, K. T., Li, R., and Sudjianto, A. (2006) Design and modeling for computer experiments. Chapman \& Hall/CRC, Boca Raton, FL

%\item
%Fang, K. T, Lin, D. K. J., Winker, P., Zhang, Y. (2000), Uniform design: Theory and application. \textit{Technometrics}, 42, 237--248

\item
Fu, J. and Wang, L., 2002. A random-discretization based Monte Carlo sampling method and its applications. Methodology and Computing in Applied Probability 4, 5¨C25.

\item
Gelman, A., Carlin, J. B., Stern, H. S., Dunson, D. B., Vehtari, A., and Rubin, D. B., (2013), Bayesian Data Analysis, Third Edition, Chapman \& Hall/CRC, Boca Raton, FL

\item
Ghosh, J. K., Delampady, M., and Samanta, T. (2006), \textit{An Introduction to Bayesian Analysis: Theory and Methods}, Springer. New York.

\item
Haario, H., Saksman, E., and Tamminen, J. (2001), An adaptive Metropolis algorithm, Bernoulli, 7, 223--242

\item
Hastings, W. K. (1970), Monte Carlo sampling methods using Markov chains and their applications, Biometrika, 57, 97--109.

%\item
%He, X. (2017), Rotated sphere packing designs, \textit{Journal of the American Statistical Association}, 112, 1612--1622.

%\item
%He, X. (2021), Lattice-based designs possessing quasi-optimal separation distance on all projections, \textit{Biometrika}, 108, 443--454.

%\item
%Huang, Y. and Zhou, Y. (2022), Convergence of uniformity criteria and the application in numerical integration. Mathematics, 10(19), 3717.

\item
Joe, S. (2004). Construction of good rank-1 lattice rules based on the weighted star discrepancy. In Monte Carlo and Quasi-Monte Carlo Methods 2004 (pp. 181-196). Berlin, Heidelberg: Springer Berlin Heidelberg.

\item
Johnson, M., Moore, L., and Ylvisaker, D. (1990), Minimax and maximin distance design, \textit{Journal of Statistical Planning and Inference}, 26, 131--148.

\item
Joseph, V. R. (2012): Bayesian computation using design of experiments-based interpolation technique, Technometrics, 54:3, 209-225

%\item
%Joseph, V. R. (2016), Space-filling designs for computer experiments: A review, \textit{Quality Engineering}, 28, 28--35.

%\item
%Joseph, V. R., Gul, E., and Ba, S. (2015), Maximum projection designs for computer experiments, \textit{Biometrika}, 102, 371--380.

%\item{}
%Loeppky, J. L, Sacks, J., and Welch, W. J. (2009). Choosing the sample size of a computer experiment: A practical guide, \textit{Technometrics}, 51, 366--376.

\item{}
Li, C., Cui, X., and Xiong, S. (2023), Design and analysis of computer experiments with both numeral and distributional inputs, Technometrics, 65, 406-417.

\item{}
Liu, J. S. (2001). Monte Carlo Strategies in Scientific Computing. Springer, New York.

\item{}
Mak, S. and Joseph, V. R. (2018), Support points, The Annals of Statistics, 46, 2562--2592.

%\item{}
%Maronna, R.A., Martin, R.D., Yohai, V.J., (2006) Robust Statistics: Theory and Methods, Wiley, New York.

\item{}
Massart, P. (2006). Concentration Inequalities and Model Selection. Springer, New York.

%\item{}
%Mckay, M. D., Beckman, R. J., and Conover, W. J. (1979), A comparison of three methods for selecting values of input variables in the analysis of output from a computer code, %\textit{Technometrics}, 21, 239--245.

\item{}
Metropolis, N., Rosenbluth, A.W., Rosenbluth, M. N., Teller, A. H., and Teller, E. (1953), Equation of state calculation by fast computing machines,
Journal of Chemical Physics, 21, 1087--1092.

\item{}
Minka, P. (2001), Expectation propagation for approximate Bayesian inference, Uncertainty in Artificial Intelligence, 17, 362--369.

\item{}
Niederreiter, H. (1992), \textit{Random Number Generation and Quasi-Monte Carlo Methods}, SIAM, Philadelphia.

%\item{}
%O'Hagan, A., Murphy, T.B., Gormley, I.C., (2012). Computational aspects of fitting mixture models via the expectation-maximization algorithm. Computational
%Statistics and Data Analysis 56, 3843--3864.

\item{}
Park, J. S. (2001). Optimal Latin-hypercube designs for computer experiments. \textit{J. Statist. Plann. Inference}, 39, 95--111.

\item{}
Park, T. and Casella, G. (2008) The Bayesian Lasso, Journal of the American Statistical Association, 103, 681--686.

\item{}
Rahimi, A. and Recht, B. (2007) Random features for large-scale kernel machines, Advances in Neural Information Processing Systems, 20.

\item{}
Rasmussen, C. E. and Williams, C. K. I. (2006). \textit{Gaussian Processes for Machine Learning}, MIT Press.

\item{}
Robert, C., and Casella, G. (2004), Monte Carlo Statistical Methods, New York: Springer-Verlag

\item{}
Rue, H., Martino, S., and Chopin, N. (2009), Approximate Bayesian inference for latent Gaussian models by using integrated nested Laplace approximations (with discussion), Journal of the Royal Statistical Society, Series B, 71, 319--392.

\item{}
Santner, T. J., Williams, B. J., and Notz, W. I. (2018). \textit{The Design and Analysis of Computer Experiments}, 2nd Edition, Springer. New York.

\item{}
Shao, J. (1999). \textit{Mathematical Statistics}, Springer. New York.

\item{}
Tibshirani, R., 1996, Regression shrinkage and selection via the lasso. Journal of the Royal Statistical Society, Series B, 58, 267-288.

%\item{}
%Tierney, L., and Kadane, J. B. (1986), Accurate Approximations for Posterior Moments and Marginal Densities, Journal of the American Statistical
%Association, 81, 82-86.

\item{}
Wainwright, M. J. (2019) High-Dimensional Statistics: A Non-Asymptotic Viewpoint, Cambridge University Press.

\item{}
Wang, L. and Lee, C. H. (2014), Discretization-based direct random sample generation, \textit{Computational Statistics and Data Analysis}, 71, 1001--1010.

\item{}
Wu, Y., Xiong, S., and Chien, P. (2026), Random Fourier features based Gaussian process models for emulation of stochastic simulations, SIAM/ASA Journal of Uncertainty Quantification, 14, 48-76.

\item{}
Yi, S.Y., Liu, Z., Liu, M. Q., and, Zhou Y.D. (2023). Global likelihood sampler for multimodal distributions, Journal of Computational and Graphical Statistics, 32: 927-937

\item{}
Zhao, J., Liu, X., Du, B., and Liu, Y. (2025), Approximation error from discretizations and its applications. The Annals of Statistics, 53(2), 589-614.

\item{}
Zimmerman, D. L. and Cressie, N. (1992), Mean squared prediction error in the spatial linear model with estimated covariance parameters, Annals of the Institute of Statistical Mathematics, 44, 27-43.

\item{}
Zou, H., (2006), The adaptive Lasso and its oracle properties. Journal of the American Statistical Association, 101, 1418-1429.

\end{description}}

\end{document}